\documentclass{llncs}
\usepackage{graphicx}
\usepackage{amsmath}
\usepackage{amssymb}
\usepackage{amsmath}
\usepackage{amssymb}
\usepackage{algorithm}

\newcommand{\be}{\begin{equation}}
\newcommand{\ee}{\end{equation}}
\newcommand{\nin}{\noindent}
\newcommand{\pF}{{\Pr}_{func}}
\newcommand{\pf}{{\Pr}_{perm}}

\numberwithin{equation}{section}
\numberwithin{lemma}{section}
\numberwithin{proposition}{section}

\begin{document}

\title{How many queries are needed to distinguish a truncated random permutation from a random function?}

\author{Shoni Gilboa \inst {1}, Shay Gueron \inst {2, 3} \and Ben Morris \inst{4}}
\institute{
The Open University of Israel, Raanana 43107 , Israel
\and
University of Haifa, Haifa 31905, Israel, ${ }^{3}$  Intel Corporation
\newline
${ }^{4}$ UC Davis
}

\maketitle
\centerline{ \today }
 \begin{abstract}

An oracle chooses a function $f$ from the set of $n$ bits strings to itself, which is either a randomly chosen permutation or a randomly chosen function. When queried by an $n$-bit string $w$, the oracle computes $f(w)$, truncates the $m$ last bits, and returns only the first $n-m$ bits of $f(w)$. How many queries does a querying adversary need to submit in order to distinguish the truncated permutation from the (truncated) function? \\
In 1998, Hall et al. \cite{Hall} showed an algorithm for determining (with high probability) whether or not $f$ is a permutation, using $O(2^{\frac{m+n}{2}})$ queries. They also showed that if $m < n/7$, a smaller number of queries will not suffice. For $m > n/7$, their method gives a weaker bound.
In this note, we first show how a modification of the approximation method used by Hall et al. can solve the problem completely. It extends the result to practically any $m$, showing that $\Omega(2^{\frac{m+n}{2}})$ queries are needed to get a non-negligible distinguishing advantage. However, more surprisingly, a better bound for the distinguishing advantage can be obtained from a result of Stam \cite{Stam} published, in a different context, already in 1978. We also show that, at least in some cases, the bound in \cite{Stam} is tight. 
\end{abstract}

{\small
\begin{quote}
\textbf{Keywords:} Pseudo random permutations, pseudo random functions, advantage.
\end{quote}}

\section{Introduction}

Distinguishing a randomly chosen permutation from a random function is a combinatorial problem which is fundamental in cryptology. A few examples where this problem plays an important role are the security analysis of block ciphers, hash and MAC schemes.
 
One formulation of this problem is the following. An oracle chooses a function $F: \{0, 1\}^n \to \{0, 1\}^n$, which is either a randomly (uniformly) chosen permutation of $\{0, 1\}^n$, or a randomly (uniformly) chosen function from $\{0, 1\}^n$ to $\{0, 1\}^n$. An adversary selects a ``querying and guessing'' algorithm. He first uses it to submit $q$ (adaptive) queries to the oracle, and the oracle responds with $F(w)$ to the query $w \in \{0, 1\}^n$. After collecting the $q$ responses, the adversary uses his algorithm to guess whether or not $F$ is a permutation.  The quality of such an algorithm (in the cryptographic context) is the ability to distinguish between the two cases (rather than successfully guessing which one it is). It is measured by the difference between the probability that the algorithm outputs a certain answer, given that the oracle chose a permutation, and the probability that the algorithm outputs the same answer, given that the oracle chose a function. This difference is called the "advantage" of the algorithm. We are interested in estimating \emph{Adv}, which is the maximal advantage of the adversary, over all possible algorithms, as a function of a budget of $q$ queries. 

The well known answer to this problem is based on the simple ``collision test'' and the Birthday Problem:
$$Adv=1-\left(1-\frac{1}{2^n}\right)\left(1-\frac{2}{2^n}\right)\ldots\left(1-\frac{q-1}{2^n}\right).$$
Since for every $1\leq k\leq q-1$
$$1-\frac{q}{2^n}\leq\left(1-\frac{k}{2^n}\right)\left(1-\frac{q-k}{2^n}\right)\leq\left(1-\frac{q}{2^{n+1}}\right)^2,$$
we get, for $q\leq 2^n$, that
\begin{equation}\label{birthday}1-e^{-\frac{q(q-1)}{2^{n+1}}}\leq 1-\left(1-\frac{q}{2^{n+1}}\right)^{q-1}\leq Adv \leq 1-\left(1-\frac{q}{2^n}\right)^{\frac{q-1}{2}}\leq\frac{q(q-1)}{2^{n+1}}.\end{equation}
This result implies that the number of queries required to distinguish a random permutation from a random function, with success probability significantly larger than, say, $\frac{1}{2}$, is $\Theta(2^{\frac{n}{2}})$.
We now consider the following generalization of this problem: \\

\nin {\bf Problem 1. [Distinguishing a truncated permutation]}
Let $0 \le m<n$ be integers. An oracle chooses $c \in \{0, 1\}$. If $c = 1$, it picks a permutation $p$ of $\{0, 1\}^n$ uniformly at random, and if $c = 0$, it picks a function $f: \{0, 1\}^n \to \{0, 1\}^n$ uniformly at random. 
An adversary is allowed to submit queries $w \in \{0, 1\}^n$ to the oracle. The oracle computes $\alpha = p(w)$ (if $c=1$) or $\alpha = f(w)$ (if $c=0$), truncates (with no loss of generality) the last $m$ bits from $\alpha$, and replies with the remaining $(n-m)$ bits. The adversary has a budget of $q$ (adaptive) queries, and after exhausting this budget, is expected to guess $c$. 
{\it How many queries does the adversary need in order to gain non-negligible advantage?} \\ Specifically, we seek $q_{\frac{1}{2}}=\min\{q\mid Adv\geq\frac{1}{2}\}$ as a function of $m$ and $n$.

\section{So, how many queries are really needed?}

\subsubsection{The Birthday bound (folklore):}
We start with remarking that the classical "Birthday" bound $q_{\frac{1}{2}}=\Omega(2^{n/2})$ is obviously valid as a bound for the adversary's advantage in Problem 1. 
In fact, any algorithm that the adversary can use with the truncated replies of $(n-m)$ bits from $f(w)$ can also be used by the adversary who sees the full $f(w)$ (he can ignore $m$ bits and apply the same algorithm). \newline

Of course, we are looking for a better upper bound that would reflect the fact that the adversary receives less information when $f(w)$ is truncated. 
We have the following bounds for Problem 1.

\subsubsection{Hall et al. \cite{Hall} (1988):}
Problem 1 was studied by Hall et al. \cite{Hall} in 1998. The authors showed an algorithm that gives a non-negligible distinguishing advantage using $q = O (2^{(n+m)/2})$ queries (for any $m$). They also proved the following upper bound:
\begin{equation}\label{Hall_result}
Adv \le 5\left(\frac{q}{2^{\frac{n+m}{2}}}\right)^{\frac{2}{3}}+\frac{1}{2}\left(\frac{q}{2^{\frac{n+m}{2}}}\right)^3\frac{1}{ 2^{\frac{n-7m}{2}}}
\end{equation}

For $m\leq n/7$ the bound in (\ref{Hall_result}) implies that $q_{\frac{1}{2}}=\Omega(2^{\frac{m+n}{2}})$. However, for larger values of $m$, the bound on $q_{\frac{1}{2}}$that is offered by (\ref{Hall_result}) deteriorates, and becomes (already for $m>n/4$)worse than the trivial "Birthday" bound $q_{\frac{1}{2}}=\Omega(2^{n/2})$. 

Hall et al. \cite{Hall} conjectured that $\Omega(2^{\frac{m+n}{2}})$ queries are needed in order to get a non-negligible advantage, in the general case. 

\subsubsection{Bellare and Impagliazzo \cite{BI} (1999):}

Theorem 4.2  in \cite{BI} states that 
\begin{equation}\label{eq:BI}Adv=O(n)\frac{q}{2^{\frac{n+m}{2}}}\end{equation}
whenever $2^{n-m}<q<2^{\frac{n+m}{2}}$. 

This implies that $q_{\frac{1}{2}}=\Omega(\frac{1}{n}2^{\frac{m+n}{2}})$ for $m>\frac{1}{3}n+\frac{2}{3}\log_2 n+\Omega(1)$.

\subsubsection{Gilboa and Gueron \cite{GG} (2013):}

The method used to show \eqref{Hall_result} can be pushed to prove the conjecture in \cite{Hall} for (almost) every $m$. 
In particular, it can be shown that if $m\leq n/3$ then
\begin{equation}\label{thm1}Adv\leq 2\sqrt[3]{2}\left(\frac{q}{2^{\frac{n+m}{2}}}\right)^{\frac{2}{3}}+\frac{2\sqrt{2}}{\sqrt{3}}\left(\frac{q}{2^{\frac{n+m}{2}}}\right)^{\frac{3}{2}}+\left(\frac{q}{2^{\frac{n+m}{2}}}\right)^2,\end{equation}
and if \(\frac{n}{3}<m\leq n-4-\log_2 n\) then
\begin{equation}\label{thm2}Adv\leq 3\left(\frac{q}{2^{\frac{n+m}{2}}}\right)^{\frac{2}{3}}+2\left(\frac{q}{2^{\frac{n+m}{2}}}\right)+5\left(\frac{q}{2^{\frac{n+m}{2}}}\right)^2+\frac{1}{2}\left(\frac{2q}{2^{\frac{n+m}{2}}}\right)^{\frac{n}{n-m}}.\end{equation}

This implies that 
$q_{\frac{1}{2}}=\Omega(2^{\frac{m+n}{2}})$ for any $0 \ge m \ge n-4-\log_2 (n)$.

\subsubsection{Stam \cite{Stam} (1978):}
Surprisingly, it turns out that Problem 1 was solved $20$ years before Hall et al. \cite{Hall}, in a different context. The bound
\begin{equation}\label{eq:Stam}Adv\leq\frac{1}{2}\sqrt{\frac{(2^{n-m}-1)q(q-1)}{(2^n-1)(2^n-(q-1)}}\leq\frac{1}{2\sqrt{1-\frac{q-1}{2^n}}}\cdot\frac{q}{2^{\frac{n+m}{2}}},\end{equation}
which is valid for all $0\leq m<n$, follows directly from a result of Stam \cite[Theorem 2.3]{Stam}. 
(Note that if $q \le \frac{3}{4} 2^{n}$ then (\ref{eq:Stam}) can be simplified to 
$Adv \le \frac{q} {  2^{ \frac{m+n}{2} } }$).

This implies that 
$q_{\frac{1}{2}}=\Omega(2^{\frac{m+n}{2}})$ for any $0 \ge m \ge n$, confirming the conjecture of \cite{Hall} in all generality ($20$ years before the conjecture was raised). 

\begin{remark}
The bound \eqref{eq:Stam} is tighter than all the bounds mentioned above, with one exception: the elementary upper bound \eqref{birthday} is better than \eqref{eq:Stam} for $q\leq 2^{\frac{n-m}{2}}$. 
\end{remark}

\section{Different methods give different bounds}

It is interesting to see how different approaches yield different bounds for Problem 1. To this end, we first define some notations.

For fixed \(m<n\) and \(q\leq 2^n\) denote \(\Omega_q :=\left(\{0,1\}^{n-m}\right)^q\). We view $\Omega_q$ as the set of all possible sequences of replies that can be given by the oracle (to the adversary's \(q\) queries).

\noindent For any \(j\geq 2\) , \(\omega\in\Omega\) let
$$
col_j(\omega)=\#\{1\leq i_1<i_2<\ldots<i_j\leq q\mid\omega_{i_1}=\omega_{i_2}=\ldots=\omega_{i_j}\}
$$

\noindent 
For $\omega=(w_1,w_2,\ldots,w_q)\in\Omega$ and $1\leq r\leq q$, let $$V_r(\omega):=\{(x_1,x_2,\ldots,x_q)\in \Omega\mid \forall 1\leq i\leq r:\,x_i=w_i\}$$
be the set of sequences of replies that are the same as $\omega$ up to the $r$-th query.

For \(\omega\in\Omega\) let \(\pf(\omega)\) and \(\pF(\omega)\) be the probabilities that \(\omega\) is the actual sequence of replies that the oracle gives to the adversary's \(q\) queries, in the case the oracle chose a random permutation or a random function, respectively.

For $1\leq r\leq q$, let 
$$Q^{(r)}_{perm}(\omega)=\frac{\pf(V_r(\omega))}{\pf(V_{r-1}(\omega))},\quad Q^{(r)}_{func}(\omega)=\frac{\pF(V_r(\omega))}{\pF(V_{r-1}(\omega))}.$$

\noindent 
Note that 
$$\pf(\omega)=\prod_{r=1}^q Q^{(r)}_{perm}(\omega),\quad \pF(\omega)=\prod_{r=1}^q Q^{(r)}_{func}(\omega).$$

\subsection{The proof method of Hall et al.}
The proof of \eqref{Hall_result} uses the general bound 
\begin{gather}\label{Adv_S:Hall}
Adv\leq\max_{\omega\in S}\left\lvert\frac{\pf(\{\omega\})}{\pF(\{\omega\})}-1\right\rvert+ \max \left\{ \pF (\overline{S}),  \pf (\overline{S}) \right\}\leq\nonumber\\
\leq 2\max_{\omega\in S}\left\lvert\frac{\pf(\{\omega\})}{\pF(\{\omega\})}-1\right\rvert+\pF (\overline{S}).\end{gather}
that holds for any $S\subseteq \Omega$.
It is applied to the set
$$S:= \left\{\omega\in\Omega:\left\lvert col_2(\omega)-\binom{q}{2}\frac{1}{2^{n-m}}\right\rvert\leq c_1\frac{q}{2^{\frac{n-m}{2}}}\; ,\; col_3(\omega)=0\right\},$$ \\

The expression 
 $\max_{\omega\in S}\left\lvert\frac{\pf(\{\omega\})}{\pF(\{\omega\})}-1\right\rvert$ is bounded by direct computations. The expression $\pF (\overline{S})$ is bounded by combining the Union Bound and the Chebyshev inequality. Finally, $c_1$ is chosen to minimize the resulting bounds.

\subsection{The proof method of Gilboa and Gueron}

To get \eqref{thm1} (for $m\leq n/3$), we can apply the slightly better (than \eqref{Adv_S:Hall}) bound
\begin{gather}\label{eq:Adv_S}
Adv\leq\frac{1}{2}\max_{\omega\in S}\left\lvert\frac{\pf(\{\omega\})}{\pF(\{\omega\})}-1\right\rvert+\frac{1}{2}\left(\pF (\overline{S})+\pf (\overline{S}) \right)\leq\nonumber\\
\leq\max_{\omega\in S}\left\lvert\frac{\pf(\{\omega\})}{\pF(\{\omega\})}-1\right\rvert+ \min \left\{ \pF (\overline{S}),  \pf (\overline{S}) \right\} .\end{gather}
to the set
$$S:= \left\{\omega\in\Omega:\left\lvert col_2(\omega)-\binom{q}{2}\frac{1}{2^{n-m}}\right\rvert\leq c_2\frac{q^{2/3}2^{2m/3}}{2^{n/3}}\, ,\, col_3(\omega)\leq
c_3\frac{q^{3/2}}{2^n}\right\}$$
Here, $c_2,c_3$ are chosen to minimize the bound.
Again, $\max_{\omega\in S}\left\lvert\frac{\pf(\{\omega\})}{\pF(\{\omega\})}-1\right\rvert$ is bounded by direct (elaborate) computation, and $\pF (\overline{S})$ is bounded by combining (via the Union Bound) the Chebyshev inequality and the Markov inequality.

The bound \eqref{thm2} (for $n/3<m\leq n-4-\log_2 n$) follows similarly by examining the set
$$S:= \left\{\omega\in\Omega:\left\lvert col_{j+1}(\omega)-\binom{q}{j+1}\frac{1}{2^{j(n-m)}}\right\rvert\leq\alpha_j\;\forall 1\leq j\leq t-1\, ,\, col_{t+1}(\omega)\leq\beta\right\}$$
for $t:=\left\lceil\frac{n+m}{n-m}\right\rceil$ and $\alpha_1,\ldots,\alpha_{t-1},\beta$ which are chosen to optimize the bound. 

\subsection{The proof method of Bellare and Impagliazzo}

Bellare and Impagliazzo also used \eqref{eq:Adv_S}, for the set $S$ of all $\omega\in\Omega$ satisfying (for suitable $\delta$ and $\lambda$):

\begin{enumerate}
\item For any $1\leq r\leq q$,
$$\left|\log\frac{Q^{(r)}_{perm}(\omega)}{Q^{(r)}_{func}\omega)}\right|\leq\frac{3\delta}{2}$$ 
\item  For any $1\leq r\leq q$,
$$\left|\sum_{x\in V_{r-1}(\omega)}\frac{{\pF}(x)}{{\pF}(V_{r-1}(\omega))}\log\frac{Q^{(r)}_{perm}(x)}{Q^{(r)}_{func}(x)}\right|\leq\frac{\delta^2}{2},$$ 
\item 
$$\left|\log\frac{{\pf}(\omega)}{{\pF}\omega)}-\sum_{r=1}^q\sum_{x\in V_{r-1}(\omega)}\frac{{\pF}(x)}{{\pF}(V_{r-1}(\omega))}\log\frac{Q^{(r)}_{perm}(x)}{Q^{(r)}_{func}(x)}\right|
\leq\frac{\delta(\delta+3)\lambda\sqrt{q}}{2}.
$$ 
\end{enumerate}

The expression $\pF (\overline{S})$ is bounded by combining the Azuma inequality and the observation that for any $1\leq r\leq q$,
\begin{gather*}0\geq\sum_{\omega\in\Omega}Q^{(r)}_{func}(\omega)\log\frac{Q^{(r)}_{perm}(\omega)}{Q^{(r)}_{func}(\omega)}\geq -\frac{1}{2} \left( \max_{\omega\in S}\left\lvert\frac{Q^{(r)}_{perm}(\omega)}{Q^{(r)}_{func}(\omega)}-1\right\rvert\right)^2,\end{gather*}

\subsection{The proof method of Stam}

Stam's approach observes that by Pinsker's inequality (1960) \cite{Pinsker} 
\footnote{
The inequality as used in \eqref{Pinsker} was established independently by Csisz\'ar (1967) \cite{Csiszar}, Kemperman (1968) \cite{Kemperman}, and Kullback \cite{Kullback}. Pinsker proved the inequality with a worse constant.} we have
\begin{gather}
\label{Pinsker}
Adv\leq\frac{1}{2}\sum_{\omega\in\Omega}\left|\pf(\omega)-\pF(\omega)\right|\leq\nonumber\\
\le \sqrt{\frac{1}{2}\sum_{\omega\in\Omega}\pf(\omega)\log\frac{\pf(\omega)}{\pF(\omega)}}. \end{gather}

He then uses the decomposition 
\begin{gather*}
\sum_{\omega\in\Omega}\pf(\omega)\log\frac{\pf(\omega)}{\pF(\omega)}=\sum_{r=1}^q\sum_{\omega\in\Omega}\pf(V_{r-1}(\omega))Q^{(r)}_{perm}(\omega)\log\frac{Q^{(r)}_{perm}(\omega)}{Q^{(r)}_{func}(\omega)}, 
\end{gather*}
direct (exact) computations, and the concavity of the $\log$ function. 

\section{Stam's bound is sometimes sharp}\label{sec:Stam_sharp}
In the case $m=n-1$ (i.e., the oracle returns only $1$ bit), \eqref{eq:Stam} gives 
\begin{equation*}
Adv\leq\frac{1}{2}\sqrt{\frac{q(q-1)}{(2^n-1)(2^n-(q-1)}}\leq\frac{1}{\sqrt{2-\frac{q-1}{2^{n-1}}}}\cdot\frac{q}{2^n}.
\end{equation*}
In this section we show that this bound is essentially sharp. 

With no loss of generality we may assume $q$ is even and $q\leq\frac{1}{2}2^n$. We define the following adversarial algorithm. 

\begin{itemize}
\item []
{\bf Algorithm 1.} \\ 
Collect the answers (which are, in this case, just bits) of $q$ arbitrary queries. \\ 
Compute the difference $\Delta$ between the number of $0$'s and $1$'s. \\
If  $\Delta \le \sqrt{q}/2$, guess that the oracle was using a truncated random permutation. Otherwise, guess that the oracle was using a random function.
\end{itemize}

The advantage of Algorithm 1 is
\begin{gather*}
\sum_{|k-(q-k)|<\sqrt{q}/2}\binom{q}{k}\left(\frac{\prod_{i=1}^k(2^{n-1}-(i-1))\cdot\prod_{i=1}^{q-k}(2^{n-1}-(i-1))}{\prod_{i=1}^q(2^n-(i-1))}-\frac{1}{2^q}\right)=\\
=\sum_{|k-(q-k)|<\sqrt{q}/2}\binom{q}{k}\frac{1}{2^q}\left(\frac{\prod_{i=1}^k(2^n-2(i-1))\cdot\prod_{i=1}^{q-k}(2^n-2(i-1))}{\prod_{i=1}^q(2^n-(i-1))}-1\right)
\end{gather*}
We show that 
\begin{gather}
\binom{q}{k}\frac{1}{2^q}\geq \frac{1}{2\sqrt{q}}\label{eq:b},\\
p_k:=\frac{\prod_{i=1}^k(2^n-2(i-1))\cdot\prod_{i=1}^{q-k}(2^n-2(i-1))}{\prod_{i=1}^q(2^n-(i-1))}>1+\frac{q/2}{2^n}\label{eq:p}
\end{gather}
for any $k$ such that $|k-(q-k)|<\sqrt{q}/2$. From this, we can conclude that 
$$Adv>\sqrt{q}\frac{1}{2\sqrt{q}}\frac{q/2}{2^n}=\frac{q/4}{2^n}.$$

First, note that for $k=q/2$.
\begin{gather}
\binom{q}{q/2}\frac{1}{2^q}=\frac{1}{2}\prod_{i=2}^{q/2}\frac{2i-1}{2i}\geq\frac{1}{2}\prod_{i=2}^{q/2}\frac{\sqrt{i-1}}{\sqrt{i}}=\frac{1}{\sqrt{2q}}\label{eq:b_q/2},\\
p_{q/2}=\prod_{i=1}^{q/2}\left(1+\frac{1}{2^n-(2i-1)}\right)\geq\left(1+\frac{1}{\frac{1}{2}2^n}\right)^{q/2}\geq 1+\frac{q}{2^n}\label{eq:p_q/2}.
\end{gather}
Since for any $0\leq j<q/2$
\begin{gather*}
\frac{\binom{q}{j}}{\binom{q}{j+1}}=1-\frac{q-2j-1}{q-j}>1-\frac{2(q-2j-1)}{q},\\
\frac{p_j}{p_{j+1}}=1-\frac{2(q-2j-1)}{2^n-2j}\geq 1-\frac{4(q-2j-1)}{2^n},
\end{gather*}
we get that for any $\frac{q}{2}-\frac{\sqrt{q}}{4}\leq k<\frac{q}{2}$
\begin{gather*}
\frac{\binom{q}{k}}{\binom{q}{q/2}}=\prod_{i=k}^{\frac{q}{2}-1}\frac{\binom{q}{j}}{\binom{q}{j+1}}\geq\prod_{j=k}^{\frac{q}{2}-1}\left(1-\frac{2(q-2j-1)}{q}\right)\geq 1-\frac{2\sum_{j=k}^{\frac{q}{2}-1}(q-2j-1)}{q}=\\
=1-\frac{(q-2k)^2}{2q}\geq\frac{7}{8},\\
\frac{p_k}{p_{q/2}}=\prod_{i=k}^{\frac{q}{2}-1}\frac{p_j}{p_{j+1}}\geq\prod_{j=k}^{\frac{q}{2}-1}\left(1-\frac{4(q-2j-1)}{2^n}\right)\geq 1-\frac{4\sum_{j=k}^{\frac{q}{2}-1}(q-2j-1)}{2^n}=\\
=1-\frac{(q-2k)^2}{2^n}\geq 1-\frac{q/4}{2^n}.
\end{gather*}

Now, using \eqref{eq:b_q/2} and \eqref{eq:p_q/2} we get
\begin{gather*}
\binom{q}{k}\frac{1}{2^q}=\frac{\binom{q}{k}}{\binom{q}{q/2}}\binom{q}{q/2}\frac{1}{2^q}\geq\frac{7}{8}\cdot\frac{1}{\sqrt{2q}}>\frac{1}{2\sqrt{q}},\\
p_k=\frac{p_k}{p_{q/2}}p_{q/2}\geq\left(1-\frac{q/4}{2^n}\right)\left(1+\frac{q}{2^n}\right)>1+\frac{q/2}{2^n}.
\end{gather*}
The proof of \eqref{eq:b} and \eqref{eq:p} for $\frac{q}{2}< k\leq \frac{q}{2}+\frac{\sqrt{q}}{4}$ is similar.

\section{An open problem}

By combining \eqref{birthday}, \eqref{eq:Stam}, and the trivial bound $1$, we can conclude that the best known bound for Problem 1 is

\begin{equation}
\label{ALL}
Adv\leq\begin{cases}\frac{q(q-1)}{2^{n+1}}\quad & q< (1 + o(1))  ~ 2^{\frac{n-m}{2}}\\
\frac{1}{2}\sqrt{\frac{(2^{n-m}-1)q(q-1)}{(2^n-1)(2^n-(q-1)}}\quad & (1 + o(1))  ~ 2^{\frac{n-m}{2}}\leq q\leq (2 + o(1))  ~ 2^{\frac{n+m}{2}}\\
1 \quad& (2 + o(1)) ~ 2^{\frac{n+m}{2}}< q\end{cases}
\end{equation}
\noindent
By the lower bound in \eqref{birthday}, we know that the bound in \eqref{ALL} is essentially sharp for $m=0$. By our proof in Section \ref{sec:Stam_sharp}, we know that the bound in \eqref{ALL} is essentially sharp $m=n-1$. The natural question that remains open is whether the bound \eqref{ALL} is essentially sharp for all $0\leq m<n$.

\end{document}